          [1999/12/02 1.01 (PWD)]
\documentclass[a4paper,twocolumn]{esapub} % European paper

\usepackage{epsfig}
\usepackage{natbib}
\title{$^{56}$Co $\gamma$-rays from SN1998bu: COMPTEL upper limits}
\author{S. Pl\"uschke}
\author{R. Georgii}
\author{R. Diehl}
\author{W. Collmar}
\author{G. G. Lichti}
\author{V. Sch\"onfelder}
\affil{Max-Planck-Institut f\"ur extraterrestrische Physik, 85740 Garching,
Germany}
\author{\\H. Bloemen}
\author{W. Hermsen}
\affil{SRON, 3584 CA Utrecht, The Netherlands}
\author{K. Bennett}
\affil{Astrophysics Division, ESTEC, 2200 AG Noordwijk, The Netherlands}
\author{M. McConnell}
\author{J. Ryan}
\affil{Space Science Center, Univ. of New Hampshire, Durham, NH 03824, USA}

\begin{document}


\maketitle

\begin{abstract}
The type Ia supernova SN1998bu in M96 was observed by COMPTEL for a total of
88 days starting 17 days after the first detection of the supernova.
The accumulated effective observation time was 4.14 Msec. The COMPTEL
observations were performed in a special instrument mode improving
the low-energy sensitivity. We generated images in the 847 keV and 1238 keV
lines of $^{56}$Co, using improved point spread functions for the low-energy
mode. We do not detect SN1998bu. A spectral analysis of our data also confirms
the non-detection of the supernova. We discuss the event for which our upper limits
constrain the standard supernova models.
\end{abstract}
\section{The Observations}
On May 9.9 UT 1998 (TJD 10942.9) \cite{villi98} observed a supernova in M96
(NGC 3368), which was labelled SN1998bu. From spectrograms it was classified
as type Ia supernova \citep{ayani98}. Based on pre-explosion observations
and an estimate of the maximum blue light at $t_{\rm B,\,max}=10952.7\pm 0.5
{\rm TJD}$ \cite{meikle98} determined the explosion date to be May $2\pm 1$ UT
1998 (i.e. TJD $10935\pm 1$). \cite{hjorth97} determined the distance to M96
by Cepheid measurements to $11.3\pm 0.9\,{\rm Mpc}$.\\
Observations of supernovae in the optical and neighbouring wavelength bands
concentrate on information on the light curves from such events. The time
evolution of theses light curves could be understood as being powered by
reprocessed $\gamma$-ray line emission from freshly synthesised short-lived
radio-isotopes (e.g. decay chain of $^{56}{\rm Ni}$). However, due to the
creation of the low energy photons by secondary processes long after the
explosion itself most of the information on the initial state right after
the explosion is lost. Therefore the distinction of different supernova
models via their predicted light curves alone is uncertain. In contrast
$\gamma$-ray lines carry information from the very early phase of the
supernova and may allow a discrimination of the theoretical models
\citep{hoeflich98}. Depending on the particular type Ia supernova model
and the sensitivity of current days $\gamma$-ray instruments (OSSE \&
COMPTEL) the observations of $\gamma$-ray lines are limited to a maximum
distance of $\sim 15\,{\rm Mpc}$. Up to now only one type Ia supernova
(SN1991T) was marginally detected \citep{morris97}. According to the
distance estimate SN1998bu opened a second chance to observe low energy
$\gamma$-ray lines from a type Ia supernova.\\
The COMPTEL observations in direction of M96 started 17 days after the explosion
(TJD 10952) and covered the time until TJD 11071 (i.e. 136 days after the
SN). Due to some breaks we had in total 88 days of supernova observations,
which sums to an effective observation time of $\sim 4.14\cdot 10^6\,{\rm s}$.
To increase COMPTEL´s low energy sensitivity the telescope was switched into
an dedicated low-energy mode, decreasing the module thresholds of the
D2 detectors well below 600 keV. (For a detailed description of the COMPTEL
instrument see \cite{sch93}).
Due to the late start of the observation program COMPTEL missed the decay of
$^{56}{\rm Ni}$ ($\tau_{1/2}=8.8\,{\rm d}$). So the observations were focused
on the detection of 847 and 1238 keV lines from the secondary isotope
$^{56}{\rm Co}$ \citep{gomez98}.
\section{Imaging Analysis}
\begin{figure}[h!]
  \begin{center}
    \epsfig{figure=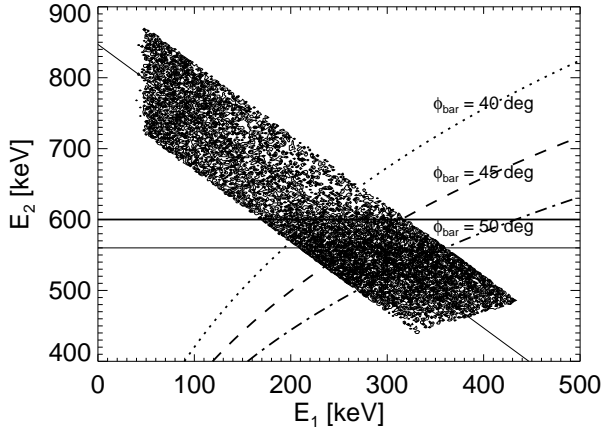,clip=,width=8cm}
    \caption{Cuts in the energy data-space in the case of the 847 keV line.}
    \label{fig:cuts}
  \end{center}
\end{figure}

The observations have been performed in a special 'low energy' mode, in which
the D2 module thresholds were lowered to the minimum.
By using appropriate point spread functions taking care of the real hardware
thresholds we gain in low energy sensitivity. Due to the strong contamination
of the D2 data due to the 511 keV background line in some D2 modules a
higher low energy cutoff (560 to 600 keV) was applied for these modules. 
However, in imaging analysis another possibility is given by an adequate
Compton scatter angle selection. Figure \ref{fig:cuts} shows the different
possibilities of 511 keV background suppression by cutting the data-space
in a $E1$-$E2$-representation (For a description of the meaurement principle
and its realisation see \cite{sch93}). The band of contour-lines represents
the  distribution of observed events with a total energy deposit (sum of
$E_1$ and $E_2$) of $847\pm75.3\,{\rm keV}$.
\\
The 511 keV contamination could be clearly seen
as an area of increased density below the horizontal 600 keV line. This
background could be suppressed either by an energy selection in $E_2$ or
an cut in the scatter angle distribution $\bar{\varphi}$.
\begin{figure*}[t!]
  \begin{center}
   \epsfig{figure=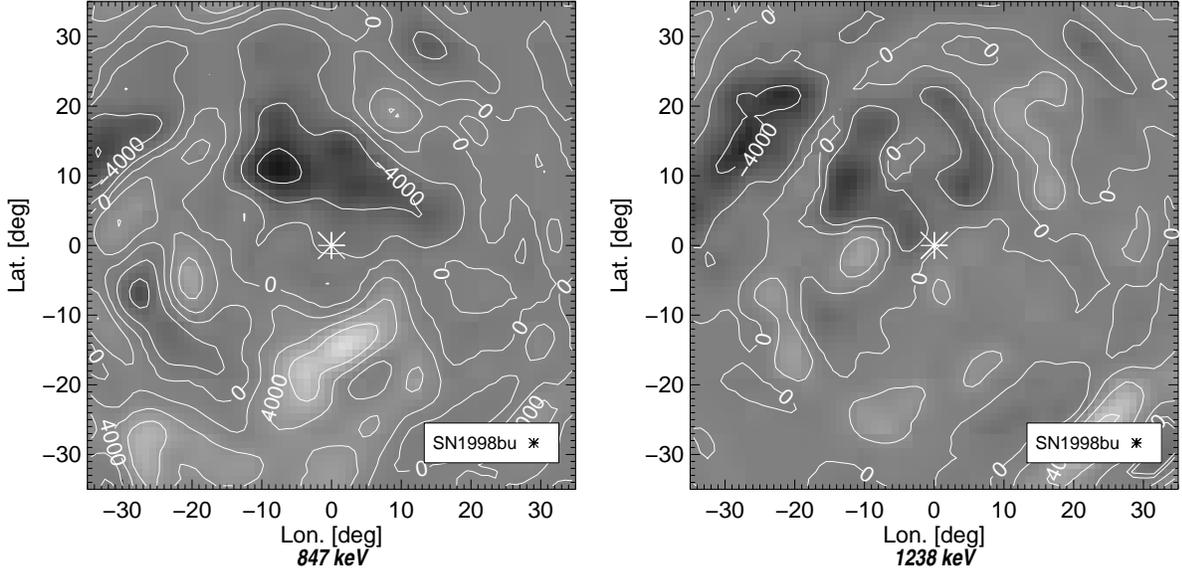,clip=,width=16cm}
    \caption{Maximum-Likelihood images (left: 847 keV/right: 1238 keV) of
     SN1998bu source region in a local coordinate system centered on the
     direction towards SN1998bu. The SN position is marked with a star.
     The contour levels give the flux in units of $10^{-8}\,{\rm ph\,
     cm^{-2}\,s^{-1}}$. 2900 in the left map correspond to $1\sigma$,
     whereas 1600 do so in the right.}
    \label{fig:maps}
  \end{center}
\end{figure*}
To really make use of the low mode data new point spread functions (PSF) have to
be generated. The COMPTEL team today uses two approaches to generate point spread
functions, one based on a Monte Carlo simulation of the detector response (SIMPSFs)
and the other based on model assumptions using measured data (MODPSFs). In standard
case with sufficient statistics both methods give rather identical point spread
functions. However, we used of both methods to generate ´low mode´ PSFs.\\
We tested the adequacy of our response treatment on Crab observations
taken in the same instrument mode. In the energy window of the low energy line
we clearly detect Crab with both PSFs. However, due to a more accurate treatment
of the detector settings in the response simulation, the image obtained with
the SIMPSF is smoother than for the MODPSF. The point-source significance rises
from $3.02\sigma$ using a standard PSF to $5.1\sigma$ by applying the new 
SIMPSF (for the MODPSF the value is $4.9\sigma$). Also in the case of the high
energy line the new PSFs increase the significance, with a smaller increase
because of the $E^{-2}$ spectrum of Crab and the smaller effect of lower
detector thresholds in the higher energy regime, the significance rises from
$0.63\sigma$ to $1.6\sigma$.\\
We applied both sets of point spread functions to the accumulated SN1998bu data
using a maximum likelihood reconstruction. Figure \ref{fig:maps} shows the images
obtained with the SIMPSFs. These images as well as the images obtained by applying
the MODPSFs show no hint of a supernova detection. Also an additional attempt using
a combined point spread function for the search of both lines simultaneously fails
in detecting SN1998bu.
We deduce $2\sigma$ upper limits: From the flux distribution of all independent
pixels of these maps their variance is determined, assuming a Gaussian distribution.
Using a Bayesian method \citep{georgii97}, which accounts for systematic as well as
statistical uncertainties, $2\sigma$ upper limits of $3.7\cdot 10^{-5}\,
{\rm ph\,cm^{-2}\,s^{-1}}$ for the 847 keV line and $1.9 \cdot 10^{-5}\,
{\rm ph\,cm^{-2}\,s^{-1}}$ for the 1238 keV line are found.
\section{Spectral Analysis}
In addition to the imaging analysis we also performed a spectral analysis
of the obtained data from the area around SN1998bu. In contrast to the
imaging analysis it is possible in the spectral domain to apply a
$\bar{\varphi}$-selection to suppress the 511 keV contamination. To achieve
an adequate suppression we applied a energy threshold to the D2 data at
600 keV. We then performed a spectral scan on a $3^{\circ}$ wide grid
around the position of SN1998bu. In each spectral analysis of this scan
we used data from a $3^{\circ}$ cone as {\it source spectrum} and data
from a $3^{\circ}\,-\,7^{\circ}$ cone mantle as {\it background spectrum}.
After fitting the background spectrum to the source spectrum the residual
spectrum should contain any excess source signal. Subsequently, we fitted
template Gaussians, with a width corresponding to the instrumental energy
resolution, to the residual spectrum to obtain the line intensities of
the expected gamma-ray lines. Using this method no significant signals
could be detected. Moreover, no significant differences between the
source position, position of SN1998bu, and the off-source positions in
scanning grid could be detected.\\
We derived $2\sigma$ upper limits by the means of the following procedure:
We generated histograms of the fitted intensities from all positions for both
lines. The width of the distributions was then interpreted as a measure of
the statistical and the systematical uncertainties of the method. In that
way we derived upper limits of $4.1\cdot 10^{-5}\,{\rm ph\,cm^{-2}\,s^{-1}}$
for the 847 keV line and $2.3\cdot 10^{-5}\,{\rm ph\,cm^{-2}\,s^{-1}}$ for
the 1238 keV line. The spectroscopically deduced upper limits are somewhat
higher because of the lower sensitivity due to the hard D2 energy cut. The
$\bar{\varphi}$-cut in imaging analysis increases the sensitivity a little
because of the $1/(E_1+E_2)$-dependence of the $\bar{\varphi}$-distribution.
In addition the background treatment in the imaging analysis is more reliable
than in the spectral domain.
\section{Interpretation and Conclusion}
For a comparison of our upper limits with recent predictions of theoretical
type Ia supernova models \citep{gomez98,kumagai98} the distance estimate
to the SN is essential. \cite{tanvir95} estimated the distance to the host
galaxy M96 by means of an HST Cepheid observation to $11.6\pm0.9\,{\rm Mpc}$,
which was revised by \cite{hjorth97} to $11.3\pm0.9\,{\rm Mpc}$.  However, 
a distance determination based on Planetary Nebulae observation suggests a
much closer distance of $9.6\pm0.6\,{\rm Mpc}$ \citep{feldm97}. This closer
distance estimate of M96 supports the deduced distance of \cite{manoz99} based
on a revised Cepheid distance calibration.
\begin{figure}[h!!!]
  %\begin{center}
    \epsfig{figure=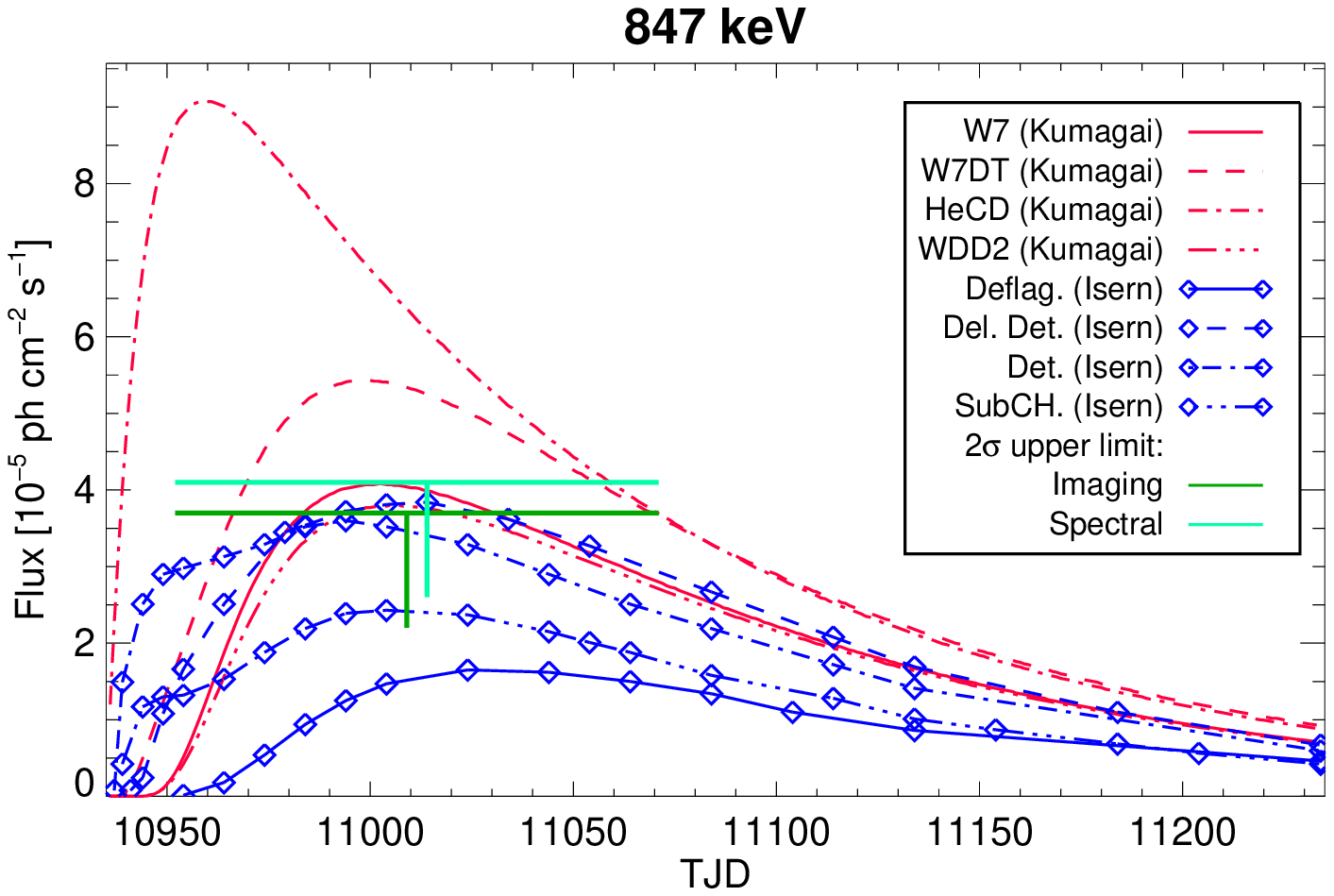,width=8.25cm}
    \epsfig{figure=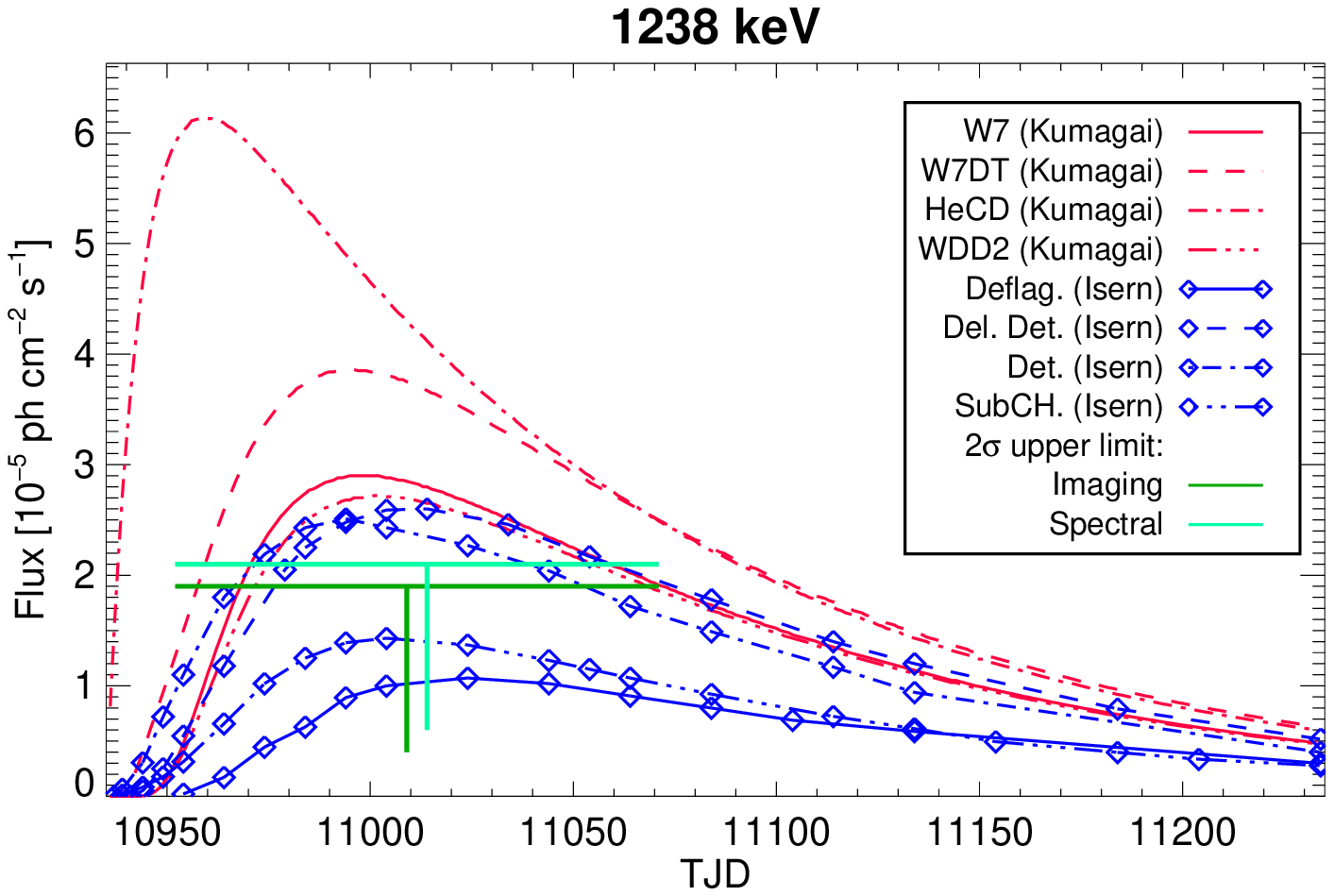,width=8.25cm}
    \caption{Predicted 847 and 1238 keV light curves from various explosion
             models. In addition the models use two different radiation transfer
             prescriptions (Gomez-Gomar et al., 1998;Kumagai, 1998). The COMPTEL
             upper limits are plotted as horizontal lines, where the length gives
             the time coverage of the COMPTEL observations.}
    \label{fig:modcomp}
  %\end{center}
\end{figure}\\
We use the greater distance estimate of 11.3 Mpc for the comparison. Figure
\ref{fig:modcomp} shows the predicted light curves from different type Ia
supernova and radiation transfer models. As pointed out by \cite{gomez98} 
the different treatments of the radiation transfer problem through the expanding
explosion shells is a major uncertainty in comparing the explosion models.
Nevertheless we exclude all but the Sub-Chandrasekhar and deflagration models.
In the case of a closer distance our results become even more constraining.\\
In summary our measurement favours the deflagration or Sub-Chandrasekhar models
for SN1998bu and renders the HeCD model as rather improbable. For detonation
models a smaller mixing than applied in tested models is needed to be compatible
with our analysis.
\vspace{-0.2cm}
\section*{Acknowledgements}
\vspace{-0.2cm}
The COMPTEL project is supported by the German 'Ministerium 
fuer Bildung und Forschung' through DLR grant 50 QV 90968.

\end{document}